# Microwave induced elastic deformation of a metallic thin film

S. B. Wang and C. T. Chan [a)]

*Department of Physics and Institute for Advanced Study, The Hong Kong*
*University of Science and Technology, Clear Water Bay, Hong Kong, P. R. China*

The microwave induced elastic deformation of a metallic thin film is computed numerically and we found that the deformation can be significantly enhanced at resonance. We show that an analytical transmission line model can reproduce the numerical results almost quantitatively and at the same time reveal the underlying physics.

Optical force has been applied broadly in recent years in small particle manipulations [1-5] and in tunable metamaterials [6]. A variety of structures such as waveguides [7,8], nanowires [9,10] and nanopatches [11-13] have been investigated, providing many functional components for opomechanical systems. Among these, the Fabry-Perot-cavity-like patch structure has been utilized in sideband cooling of a mechanical element to almost its ground state with the help of radiation pressure, making it possible to probe the quantum regime of such systems.[14,15] This shows the importance of the electromagnetic-wave-induced force in cavity optomechanics. Here we report a numerical study of microwave induced stress in a metallic-double-film cavity structure. Using a Maxwell stress tensor formulation, we determined the form and magnitude of the microwave induced deformation and we showed that the deformation can be significantly enhanced at resonance. We also showed that the numerical results can be reproduced analytically using a simple model which can help to reveal the underlying physics.

We consider a three-dimensional structure consisting of two metallic thin films clamped at the edges by a surrounding dielectric frame ($\varepsilon_r = 2.3$) as shown schematically in Fig. 1. The two metallic films have a length of *a*, width of *b* and thickness of *t* and they are separated by an air gap of $d = w - 2t$, where *w* is the thickness of the surrounding dielectric. We will consider the mechanical deformation induced by an incident plane wave at the microwave frequency. The incident field has the form: $\mathbf{E}_{inc} = \hat{x} \cdot E_0 e^{i(-k_0 z - \omega t)}$. The metal is assumed to be gold with a conductivity of $\sigma = 4.098 \times 10^7 \, \text{S/m}$.

Under the incidence of the plane wave, the two metallic films form an electromagnetic cavity [11,12]. The external field will cause a deformation of the films and we expect a stronger deformation if the field is enhanced significantly at resonance. For a single film, the resonance of the film excited by the incident microwave has a low fidelity and the field enhancement is not significant. For the double metallic films configuration shown in Fig. 1, the near-field coupling between the two metallic films will split the resonance into a so-called symmetric mode and an anti-symmetric mode. The anti-symmetric mode is a "dark" (subradiant) mode, which has strong fields as the electromagnetic energy is well confined in the cavity. In the anti-symmetric mode, anti-parallel currents are induced on the surface of the two films due to the time-varying electromagnetic flux. The oscillations of the currents produce a non-uniform charge distribution on the films with most of the charges accumulating at the edges along y-direction. As a result, we expect the electric field to be strong at the edges while the magnetic field should be strong near the middle of the films. Numerical results shown in Fig. 2 computed by COMSOL do verify such a distribution. In the numerical simulations, we set $a = b = 10$ mm, $t = 10$ μm, $w = 120$ μm and we assume a microwave with a total power of 200 mW impinging on the upper surface of the metallic film. Figures 2(d) and (e) show the electric and magnetic fields at the resonance frequency $f = 14.45$ GHz. The electric field between the two films is dominated by the z-component, which has opposite directions at the two edges since the accumulated charges have opposite signs. The magnetic field is dominated by the y-component, which is strong in the middle region. The red squares in Fig. 2(b) and (c) show the numerically computed current and charge distributions, which confirm our conclusion before. We will show by our analytical model later that the distributions are approximately sinusoidal.

The electromagnetic field induced stress is needed to calculate the elastic deformation of the metallic film. The electromagnetic total fields are first computed, and then the stress is evaluated by using the Maxwell stress tensor approach.[16] We note the Maxwell stress tensor approach should not be applied to calculating internal stresses inside a material if the fields penetrate the material.[17] Fortunately, in the microwave regime the fields inside the metallic film are negligibly small so that we do not need to worry about the internal stress. In the microwave regime, only the surface stress exerted on the metallic film is important, which can be obtained using the Maxwell stress tensor approach. Figure 3(a) shows the distribution pattern of the numerically computed surface stress exerted on the upper film. The stress takes negative values near the edges along y-direction while it is positive in the middle region. This corroborates intuitively with the charge and current distributions. At the edges the force is attractive because the accumulated charges have opposite signs. In the middle region the anti-parallel currents produce a repulsive force. This surface stress acts as a boundary load and the deformation caused by the stress is calculated by a finite element package (COMSOL) with the clamped boundary condition.[18] Figure 2 (a) shows the numerically computed maximum deformation of the upper film as a function of the exciting frequency, showing a typical resonance profile. We see that the deformation is strongly enhanced at the resonant frequency 14.45 GHz. Figure 3 (c) shows the numerically computed deformation pattern of the upper film at 14.45 GHz. The maximum deformation $u_z = 25$ nm happens at the center point of the film. We note that when the system is off-resonance (e.g. 15.0 GHz), the maximum deformation is only a few angstroms, which is orders of magnitude smaller than the resonance case. This indicates a very strong enhancement due to the field enhancement at the cavity resonance. Even at the resonance, the deformation is still small compared with the air gap between the two films. This is the reason that we can first compute the electromagnetic fields and then apply the corresponding surface stress to calculate the deformation. If a larger microwave power or a much thinner film is employed, the deformation may not be considered as a small perturbation, the electromagnetic part and the elastic part interact with each other and they have to be both considered simultaneously. We can solve the problem iteratively but for the configuration we are considering, this is not necessary.

In the following we show that this problem can be addressed analytically using a model that captures the essence of the physics at this frequency scale. We first obtain the electromagnetic wave induced stress using a transmission line model.[12] Under the condition $d \ll a$, the two metallic films can be treated approximately as a transmission line where the induced current and voltage are governed by the telegraph equations[19,20]:

$$\frac{\partial V(x)}{\partial x} + (R - i\omega L)I(x) = i\omega\mu_0 \int_{z_1}^{z_2} \hat{y} \cdot \mathbf{H}_{\text{inc}}(z)dz, \qquad (1)$$

$$\frac{\partial I(x)}{\partial x} + (G - i\omega C)V(x) = i\omega(G + C)\int_{z_1}^{z_2} \hat{z} \cdot \mathbf{E}_{\text{inc}}(z)dz, \qquad (2)$$

where $\mathbf{H}_{inc}(z)$ and $\mathbf{E}_{inc}(z)$ are the incident magnetic and electric fields, $R$ is the resistance per unit length, $G$ is the conductance per unit length of the medium sandwiched between the films (zero in this case), $L = \mu_0 d/b$ and $C = \varepsilon_0 b/d$ are the per-unit-length inductance and capacitance of the system, respectively. By applying the resonance boundary condition $I(0) = I(a) = 0$, we can solve the above two equations to obtain the current distribution as $I(x) = (B/A^2)[1-\cos(Ax) - \tan(Aa/2)\sin(Ax)]$, where $A = \sqrt{\omega^2 CL + i\omega RC}$ and $B = -\mu_0 \omega^2 C \int_{z_1}^{z_2} \hat{y} \cdot \mathbf{H}_{inc}(z) dz$. At the microwave frequency we take the limit $R \to 0$ and then the current distribution becomes $|I(x)| \approx |(B/A^2)\tan(Aa/2)\sin(Ax)|$ (as the tangent term is dominant at resonance) and so the current distribution on the metallic film is approximately sinusoidal. The lowest order resonance frequency is determined by $Aa/2 = \pi/2$, which gives $\omega_0 \approx \pi/(a\sqrt{LC})$. This resonance frequency is valid under the condition $d/a \to 0$. For a finite value of $d/a$ the electric field leakage effect will contribute to an additional capacitance. We take this into account by assigning an effective length to the total capacitance [21], i.e. replace $Ca$ by $Ca_{\text{eff}}$ with $a_{\text{eff}} = a(1+\alpha d/a)$, where $\alpha$ is the coefficient of the first order correction and this correction changes the resonance frequency to $\omega_0 \approx \pi/\left[a\sqrt{LC(1+\alpha d/a)}\right]$. By fitting the numerically computed current distribution in Fig. 2(b) we obtain $\alpha = 0.73$. This $\alpha$ is the only parameter in the model and the value is then used throughout our model to calculate the charge distribution and the electromagnetic surface stress. The lines in Fig. 2(b) and (c) show the analytical results of current and charge distributions, where the charge distribution is obtained by applying the conservation law of $\nabla \cdot \mathbf{j} + \partial \rho / \partial t = 0$. The analytical results agree well with numerical results obtained using full wave electromagnetic field solvers. The electromagnetic surface stress is then obtained by calculating the interaction force point by point between the charges and currents with the Coulomb's law and the Biot-Savart law. Figure 3(b) shows the analytically calculated stress distribution, where a good match with the numerical result obtained with Maxwell stress tensor approach is clear.

The elastic deformation of the film is then obtained using elastic theory. The deformation of a thin film under a vertical surface stress can be determined by the equilibrium equation [22]: $D\nabla^4 u_z - P(x,y) = 0$, where $D = Y_g t^3/\left[12(1-\nu^3)\right]$ is the flexural rigidity, $P(x,y)$ is the surface stress/pressure obtained using the transmission line model and $u_z(x,y)$ is the deformation along the z-direction. The Young's modulus and the Poisson's ration of gold are $Y_g = 79$ GPa and $\nu = 0.44$ respectively; The equation is solved by applying the clamped boundary condition: $u_z = 0$, $\partial u_z / \partial x = 0$ at $x = 0$, a; $u_z = 0$, $\partial u_z / \partial y = 0$ at $y = 0$, b. In compliance with this boundary condition, we write the trial solution as [18] $u_z = \sum_{m=1}^{\infty} \sum_{n=1}^{\infty} C_{mn} [1 - \cos(2m\pi x / a)][1 - \cos(2n\pi y / b)]$. The coefficients $C_{mn}$ are determined variationally by the Ritz method: $\partial(U - W) / \partial C_{mn} = 0$, where the strain energy $U$ is $U = D/2 \iint (\partial^2 u_z / \partial x^2 + \partial^2 u_z / \partial y^2)^2 dxdy$ and the work done by the surface stress is $W = \iint P(x, y) u_z dxdy$. By truncating the expansion orders at finite values ($m_{max} = 3$ and $n_{max} = 3$) we obtain a set of linear equations that can be solved for $C_{mn}$. Figure 3(d) shows the analytical results of the deformation, where an excellent match with the numerical results obtained using brute force simulations is again evident.

In conclusion, we have numerically and analytically studied the elastic deformation of a metallic thin film induced by an electromagnetic plane wave at the microwave frequency. We obtained the deformation pattern and showed that the deformation is greatly enhanced at resonance. Deformation on the nanometer scale (maximum value of about 25 nm) is found when utilizing a microwave with 200mW power incident on the metallic surface. A transmission line model is applied to capture the underlying physics in the electromagnetic part, which almost reproduces the numerical results of electromagnetic surface stress with a fitting fringe effect parameter. A variational method combined with Fourier expansions is then used to calculate the deformation of the metallic thin film, which gives results that match the numerical results very well. We note the high fidelity resonance employed here (derived from a dark mode due to near field coupling) is different from the Fabry-Perot resonance used in the sideband cooling of mechanical modes. The resonance behavior in our considered structure is very stable and the structure forms a microwave "drum" under the modulation of the incident wave. This provides a viable way to control the mechanical freedom of optomechanical systems and may find applications in the electromagnetic wave manipulations of macro structures in the future.

We thank Prof. H. B. Chan for the suggestions and M. Yang for the help on COMSOL simulations. This work is supported by Hong Kong CRF grant HKUST2/CRF/11G and SRFI11SC07. S. B. Wang's studentship is supported by HKUST Nano Science and Technology Program.

a) Author to whom correspondence should be addressed. Electronic mail: phchan@ust.hk

**Figure Captions:**

FIG. 1 A schematic picture showing the system under consideration. Two metallic films of dimensions $a \times b \times t$ are separated by an air gap and are clamped at the edges by a dielectric frame. The wave vector of the incident plane wave is perpendicular to the surfaces of the metallic films.

FIG. 2 Resonance enhancement of electromagnetic wave induced deformation of the metallic thin film structure. (a) Maximum displacement of the upper film as a function of the exciting frequency. (b) Numerical (squares) and analytical (line) results of the current distribution along *x*-direction. (c) Numerical (squares) and analytical (line) results of the charge distribution along *x*-direction. (d) $E_z$ component of the electric total field at resonance on the symmetry plane between the two films showing that the electric field is strongest at the edges. (e) $H_y$ component of the magnetic total field at resonance on the symmetry plane between the two films showing the magnetic field is strongest near the center.

FIG. 3 Electromagnetic surface stress and elastic deformation at resonance. (a) Electromagnetic surface stress calculated numerically by Maxwell tensor approach. (b) Electromagnetic surface stress calculated analytically by the model described in the text. (c) Numerical result of the deformation along z-direction (d) Analytical result of the deformation along z-direction.

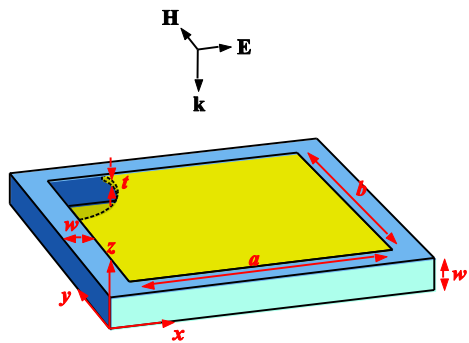

FIG. 1

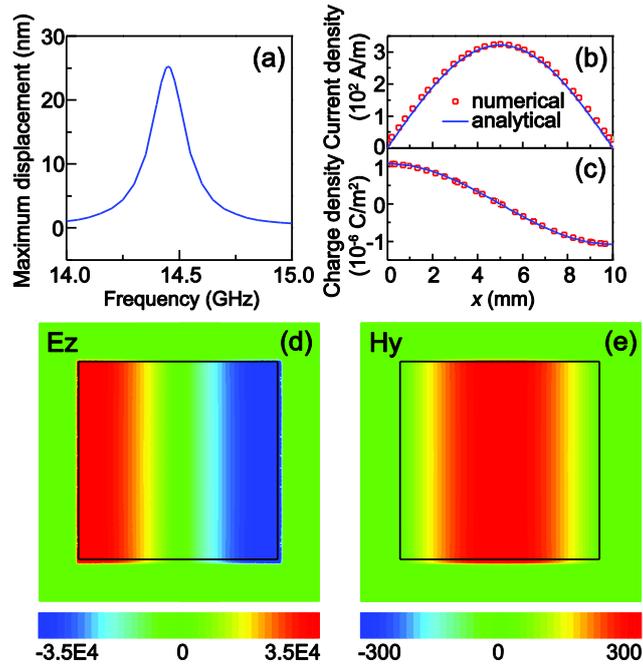

FIG. 2

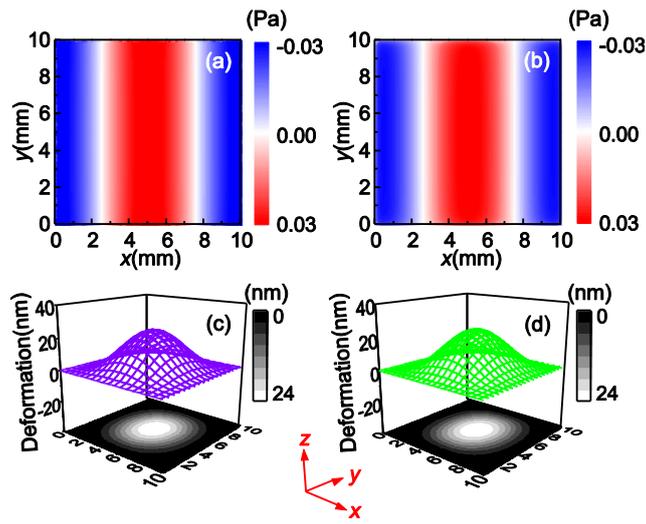

FIG. 3